\documentclass[aps,prl,twocolumn,showpacs,superscriptaddress,nofootinbib,preprintnumbers]{revtex4-2}

\usepackage{graphicx}
\usepackage{dcolumn}
\usepackage{bm}

\usepackage{style}

\begin{document}

\title{Little Red Dot Cosmology:\\ A Matter-Era Baryon Acoustic Oscillations Probe of $\Lambda$CDM}
 
\author{Jessica A. Zebrowski\,\orcidlink{0000-0003-2375-0229}} 
\email{j.z@uchicago.edu}
\affiliation{Department of Astronomy and Astrophysics, University of Chicago, 5640 South Ellis Avenue, Chicago, IL, 60637, USA}
\affiliation{Kavli Institute for Cosmological Physics, University of Chicago, Chicago, IL 60637, USA}
\affiliation{Enrico Fermi Institute, University of Chicago, 5640 South Ellis Avenue, Chicago, IL, 60637, USA}
\affiliation{Fermi National Accelerator Laboratory, Batavia, Illinois 60510}

\author{Rohan P. Naidu\,\orcidlink{0000-0003-3997-5705}}
\affiliation{Institute for Astronomy, University of Hawai'i, Honolulu HI 96822, USA}

\date{\today}

\begin{abstract}
The discovery of Little Red Dots (LRDs) with JWST provides a new source population for cosmology in a largely unmapped period of cosmic history. LRDs are most numerous at $4\lesssim z\lesssim9$, thereby bridging existing low-redshift large-scale-structure measurements and the cosmic microwave background. In this epoch, the Universe is deep in the matter-dominated era, where dark energy is dynamically negligible and the expansion history is tightly predicted in $\Lambda$CDM, making these redshifts a clean test of the standard cosmological model. Recent measurements indicate that LRDs have number densities of $\bar n\sim10^{-4}$ $h^3$ Mpc$^{-3}$, a bias rising from $b\sim 3$ to $\sim8$ over this interval, and distinctive spectrophotometric signatures -- a remarkably favorable combination for baryon acoustic oscillation (BAO) measurements. Using a Fisher forecast with these fiducial inputs, we find that a wide-field spectroscopic survey of LRDs over a
DESI-like (14,000 deg$^2$) footprint delivers a percent-level isotropic distance scale measurement of $D_V/r_d$
across four bins spanning $4\lesssim z\lesssim9$. LRDs are therefore a compelling target for future wide-field near- to mid-infrared spectroscopy aimed at mapping large-scale structure deep into the matter-dominated era.
\end{abstract}

\maketitle

\section{Introduction}
The standard $\Lambda$CDM cosmological model is pinned by measurements at the two ends of cosmic history, by the cosmic microwave background (CMB) in the early Universe, and by galaxy surveys and distance ladder measurements today. A growing set of tensions suggests the two ends might not agree. The Hubble constant inferred from the CMB under $\Lambda$CDM is in $\sim5\sigma$ tension with local distance-ladder measurements \cite{camphuis2025spt, riess2022comprehensive, riess2025perfect,freedman2025status,Planck2018,louis2025atacama}, low-redshift measurements of the clustering amplitude $S_8$ are lower than the CMB prediction \cite{Planck2018,asgari2021kids,abbott2026dark}, and DESI BAO data combined with supernovae prefer a time-evolving dark-energy equation of state over a cosmological constant at $\sim3\sigma$ \cite{DESI_DR2_II}. These discrepancies call into question whether a single set of $\Lambda$CDM parameters describes all of cosmic time, and motivate a direct measurement of the expansion history between the CMB and low-redshift large-scale-structure (LSS) tracers.

\begin{figure}[b!]
\hspace{-0.6cm}
\includegraphics[width=0.48\textwidth]{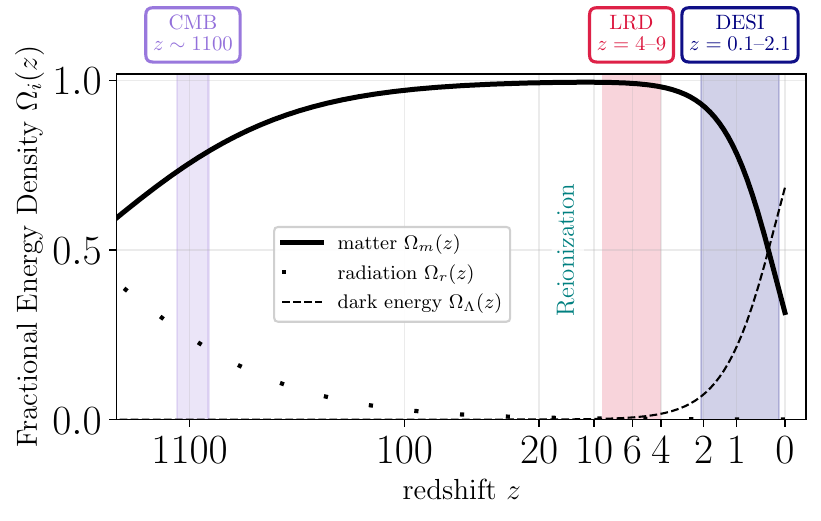}
\caption{Cosmic expansion history and current LSS tracers. The CMB
anchors the early Universe at $z\simeq1100$, while DESI galaxies map the late-time
Universe at $z\lesssim2.1$. LRDs occupy the largely unmapped interval at $4\lesssim z\lesssim9$, where the Universe is matter dominated and dark energy is dynamically negligible. A BAO measurement in this window
would provide a matter-era standard-ruler test between the CMB and the
low-redshift distance ladder.}
\label{fig:ed_hist}
\end{figure} 

A particularly clean place to make this measurement is the matter-dominated
era. As shown in Figure~\ref{fig:ed_hist}, the redshift range
$4\lesssim z\lesssim9$ lies between current low-redshift galaxy survey LSS measurements at $0.1 < z \lesssim 2.1$ and the CMB at $z \sim 1100$, at cosmic times when $\Omega_m(z)\simeq1$ and dark energy is dynamically negligible. In this epoch, the expansion history is tightly predicted in $\Lambda$CDM. 
What has been missing is a way to make this measurement. Quasars are rare, and conventional galaxy samples become challenging to survey at these redshifts. 

The discovery of Little Red Dots (LRDs; \cite[][]{Matthee2024}) with JWST may supply this missing tracer. LRDs are compact, red, broad-line sources most numerous at $4\lesssim z\lesssim9$ \cite{Kokorev24, Kocevski2024,Tanaka25, Park26, Ma26, Rinaldi26, Zhuang26, Lin26}, precisely the range needed for a matter-era standard-ruler test. Although their physical nature remains under active debate \citep[e.g.,][]{Naidu25, degraaff25cliff, degraaff25pop,  Kido25, Inayoshi25rev, Juodzbalis25, Liu25, Madau26, Sneppen26, Liu26, Chisholm26, Martins26, Zwick26, Williams26, Rantala26, Ji26, Tang26, Gentile26, Naidu26}, their relevance as a LSS tracer depends primarily on their empirical properties: their abundance, bias, and that they are easily identifiable. Current data indicate that LRDs have the promising combination of number densities greater than high-redshift quasars, a large bias that grows towards high redshift, and a combination of observable spectrophotometric signatures that is distinctive among known high-redshift galaxy and AGN populations \citep[e.g.,][]{Naidu25, degraaff25cliff, degraaff25pop}. 

In this Letter, we assess the potential of LRDs as a high-redshift LSS tracer. Due to their point-source morphology in the rest-optical \citep[][]{Furtak23, Greene24, Labbe2025, Golubchik26, Zhang25, Rinaldi25, Cloonan26} and distinctive spectroscopic signatures (e.g., broad, luminous H$\alpha$ lines; \citep[][]{Labbe2025, Kokorev2026, Matthee26, sun2026little, Davis26, Rusakov26, Scholtz26, Chen26, deugenio26}), LRDs are easy to detect and redshift. We combine current empirical estimates of their abundance and bias with a Fisher forecast for a wide-field spectroscopic survey. Using these fiducial inputs, we find that an LRD survey over a DESI-like footprint delivers percent-level isotropic baryon acoustic oscillation measurements of $D_V/r_d$ across $4\lesssim z\lesssim9$. LRDs therefore provide a concrete target for future wide-field near- to mid-infrared spectroscopy aimed at mapping large-scale structure deep into the matter-dominated era.

\section{LRDs As A Cosmological Tracer}
\begin{figure}[t!]
\hspace{-0.6cm}
\includegraphics[width=0.48\textwidth]{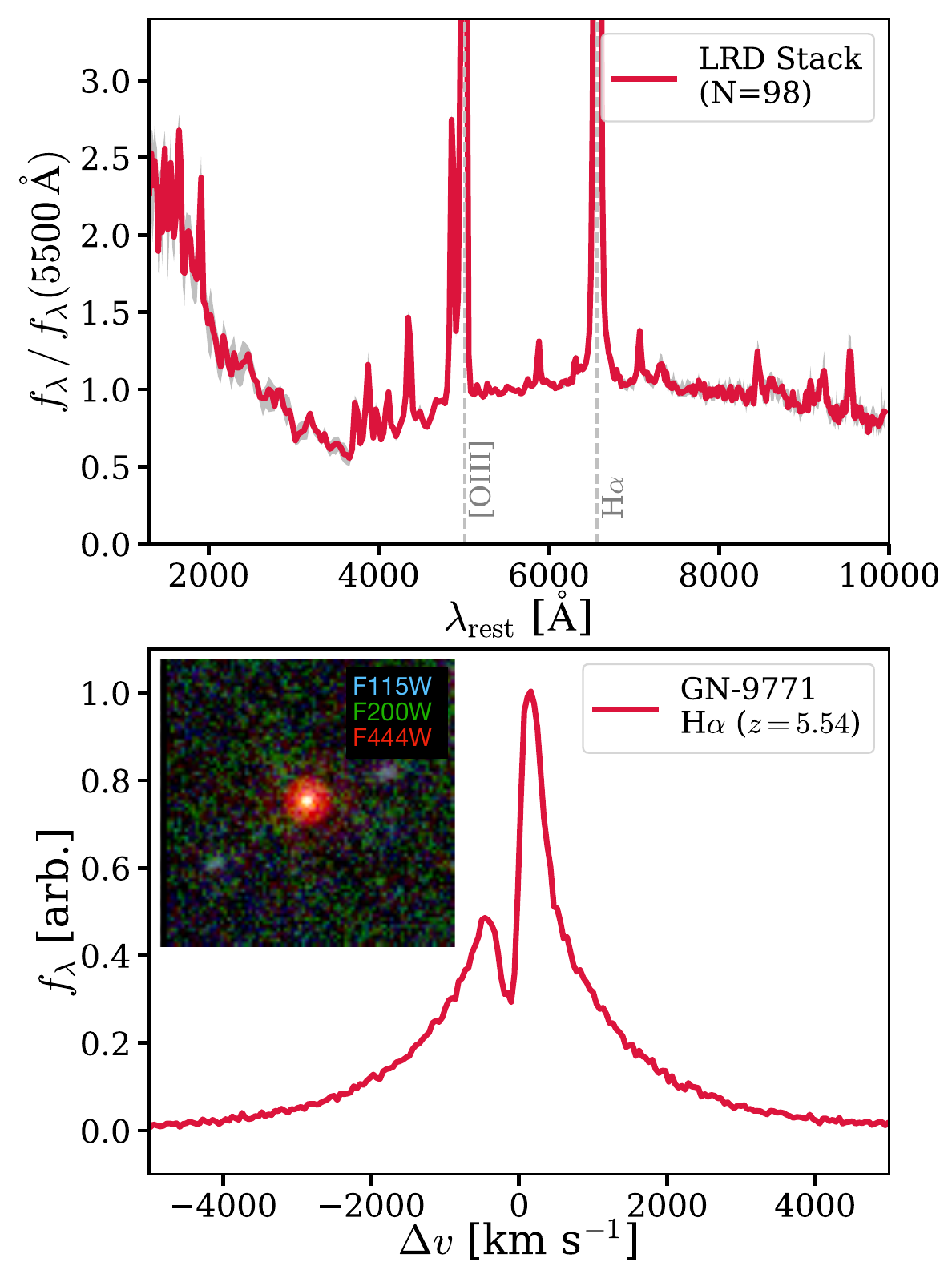}
\caption{Spectrophotometric features that make LRDs efficient redshift-survey targets. \textit{Top:} Stacked redshift-frame spectral energy distribution (SED) of 98 LRDs from \citet{sun2026little}, illustrating the characteristic ``V-shaped" continuum. \textit{Bottom:} The H$\alpha$ profile of GN-9771 at $z=5.54$ \citep[][]{Matthee2024, Torralba26, Matthee26}, showing a luminous broad emission line. \textit{Bottom Inset:} JWST/NIRCam RGB image ($2^{\prime\prime} \times 2^{\prime\prime}$) using the F115W, F200W, and F444W filters, showing the compact point-source morphology (hence ``little" and ``dot"). Any one of these features by itself is rare among high-redshift galaxies -- the trifecta (V-shaped, broad H$\alpha$, point-source) occurs only among LRDs \citep[e.g.,][]{Hviding25}.}
\label{fig:LRDstack}
\end{figure} 

\begin{figure}[t!]
\hspace{-0.6cm}
\includegraphics[width=0.48\textwidth]{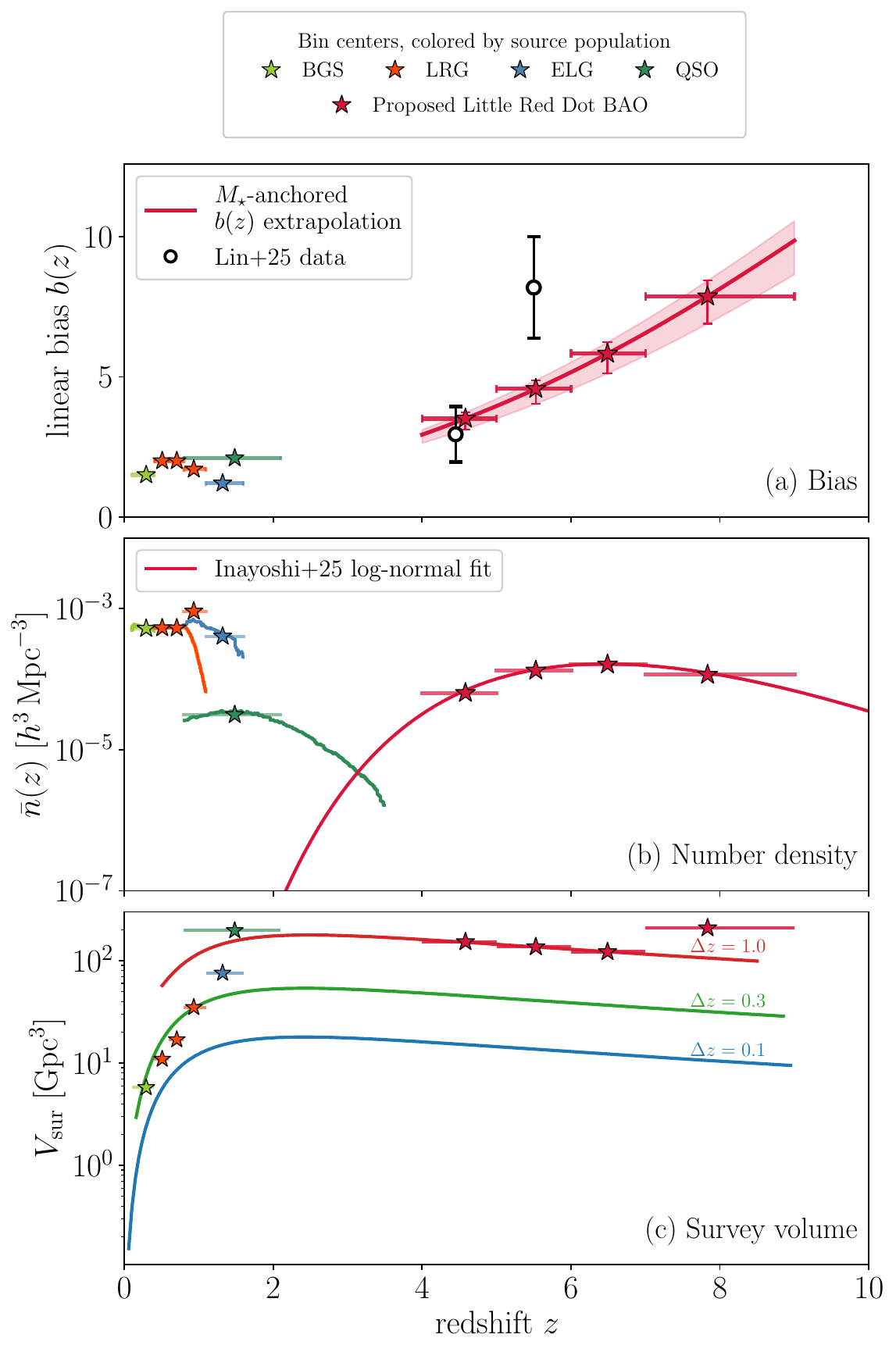}
\caption{Three inputs that make LRDs promising high-redshift BAO tracers.
Top: fiducial LRD linear bias from a stellar-mass-anchored halo model,
compared to current LRD clustering measurements. Middle: tracer number
densities, including the \citet{Inayoshi2025} LRD abundance model and DESI comparison samples. Bottom: comoving survey volume for a DESI-like 14,000 deg$^2$ footprint, shown for several redshift-bin widths. Points mark the forecast bins.}
\label{fig:threepanel}
\end{figure}

LRDs are a recent JWST discovery \citep{Matthee2024}, with $O(1000)$ LRDs now identified \citep[e.g.,][]{Kokorev24, Kocevski2024, degraaff25pop, perezgonzalez26, Weibel26bhstar, Rinaldi26}. While the physics of LRDs is far from settled, for a BAO measurement what matters is their empirical utility as an LSS tracer. A useful tracer must be identifiable in wide-field data with easy redshift assignment, abundant enough to limit shot noise, biased enough to amplify the clustering signal, and populous over a large enough redshift range to access many linear modes. Current JWST measurements suggest that LRDs satisfy all four of these conditions. We first summarize why LRDs are unusually favorable spectroscopic targets, then quantify the three inputs that control the BAO forecast: their linear bias, number density, and survey volume.

\subsection{Redshift Identification}
LRDs have three distinctive traits that make them immediately apparent in JWST data \citep[][]{Hviding25}, and unusually well suited for redshift surveys which can be seen in Figure \ref{fig:LRDstack}. First, they are point sources in the rest-optical, compact enough ($\lesssim30$ parsecs; \citep[e.g.,][]{Furtak24}) to be selected from a single imaging band covering the rest-optical. Second, they have a distinctive V-shaped spectral energy distribution, rising into the UV as well as rising into the optical, with a transition around the Balmer break region \citep[e.g.,][]{Setton25}.  Finally, they have extremely luminous, broad H$\alpha$ lines with deep absorption (FWHM$\approx1500$ km s$^{-1}$, emission line equivalent widths in the observed frame of $\approx10,000$ \AA; \citep{sun2026little, degraaff25pop, Matthee26}). Therefore, a single bright line yields a redshift identification. These distinctive features can be seen in Figure \ref{fig:LRDstack} and separate them from previously known classes of galaxies and AGN. A minimal redshift survey could deploy as few as a handful of bands of photometry \citep[e.g.,][]{Kokorev24} to detect the point-source morphology and V-shape of LRDs alongside spectroscopy to detect broad lines and return large samples with high purity and completeness. The established route to high-redshift spectroscopic samples uses deep multiband imaging for target preselection, as envisioned for the Lyman-break-galaxy and Lyman-$\alpha$-emitter samples targeted by proposed Stage-V facilities at $2 \lesssim z \lesssim 5$ \citep[e.g., Spec-S5, MegaMapper;][]{schlegel2019astro2020, schlegel2022megamapper, besuner2025spectroscopic}. Extending these approaches to still higher redshift requires progressively deeper imaging and longer-wavelength spectroscopy, while Lyman-$\alpha$ transmission is increasingly suppressed by the neutral intergalactic medium. The distinctive signatures of LRDs therefore offer an efficient path through the $z \gtrsim 4$ regime. Additionally, any LRD survey would naturally recover the brightest members of these other high-redshift populations as well.

\subsection{Bias}
Direct clustering measurements from JWST indicate that LRDs are a highly biased population \citep[][]{lin2026large}, a large advantage for a BAO tracer. For BAO constraints, the tracer power spectrum scales as $b^2 D^2(z)$ \citep[][]{kaiser1984spatial,SeoEisenstein2007,desjacques2018large}; the growth factor $D(z)$ suppresses the matter power spectrum at early times, while the bias $b$ boosts it. The halos that host LRDs ($\log_{10}(M_\star/M_\odot)=8.3^{+0.2}_{-0.4}$, $\log_{10}(M_{\rm{halo}}/M_\odot)\approx11$ at $z\approx5$; \citep[][]{sun2026little, Matthee26, lin2026large}) are unremarkable halos now, but at $z\gtrsim4$ the same halos are rare peaks of the density field and are therefore strongly biased \citep[e.g.,][]{lin2026large}. This large bias compensates for the smaller matter clustering amplitude at high redshift, leaving a clustering signal well suited to a BAO measurement.

We estimate the LRD bias in two steps. First, we estimate the redshift evolution of the LRD host halo mass from the host
stellar mass inferred by \citet{sun2026little},
$\log_{10}(M_\star/M_\odot)=8.3^{+0.2}_{-0.4}$. We hold the host stellar mass constant as a function of redshift. We invert the
\citet{behroozi2019universemachine} stellar-to-halo mass relation (SHMR) at
each redshift to obtain a fiducial halo-mass track, $M_h(z)$. This procedure
gives $\log_{10}(M_h/M_\odot)\simeq 11.03$--$10.72$ over $z=4$--$9$.

These inferred halo masses agree with the independent
estimates of \citet{lin2026large}, who find
$\log_{10}(M_h/M_\odot)=11.21^{+0.35}_{-0.32}$ at $3.9<z<5$ and
$11.04^{+0.34}_{-0.32}$ at $5<z<6$, consistent with our values over
the overlapping redshift range. We extend the calculation to $z=9$, the highest
redshift to which the \citet{behroozi2019universemachine} parameterization is
calibrated and comparable to the highest-redshift LRDs currently reported \citep[e.g.,][]{Kokorev23, Taylor25, Tripodi25, Tanaka25}.

We then convert this halo mass track to a linear bias using the
\citet{tinker2010large} halo-bias relation. The resulting bias track is shown in the top panel of Fig.~\ref{fig:threepanel}. For the four LRD forecast bins
used in this work, the inferred halo biases are $b_{\rm LRD} = 3.4,\ 4.5,\ 5.8,\ 8.1$
at bin centers $z = 4.5,\ 5.5,\ 6.5,\ 8.0$. Across all four bins the LRD bias exceeds that of every DESI tracer ($b \approx 1.2-2.1$), rising to $\sim 4\times$
the DESI samples at the highest redshift as seen in Table \ref{tab:forecast_inputs}. The corresponding clustering amplitudes are \(b(z)D(z)=0.80,\,0.89,\,0.99,\) and \(1.13\), comparable to the \(b(z)D(z)\simeq0.6-1.6\) spanned by DESI tracers. The large LRD bias compensates for the suppressed matter growth at high redshift, yielding a clustering amplitude similar to that of established BAO samples.

This procedure is intentionally conservative. Direct clustering measurements of LRDs already suggest very large bias, as shown by the open white points from \citet{lin2026large} in Fig.~\ref{fig:threepanel}, including values comparable to or larger than our fiducial track at $z>5$. Rather than fit the current sparse clustering measurements directly, we use the smooth $M_\star$-anchored bias model as a baseline forecast input. If the highest measured LRD biases persist in larger samples, the clustering signal and BAO constraining power would be stronger than in the fiducial forecasts presented here.

\subsection{Number Density}
LRDs are also sufficiently abundant for a high-redshift BAO measurement. Current measurements indicate number densities of $\approx10^{-4}$ to $10^{-5}$ cMpc$^{-3}$ at $z\approx4-8$, comparable to $L^{*}$ galaxies and $\gtrsim100\times$ more common than classical UV-bright quasars at these redshifts \citep{Matthee2024}. Their abundance falls by 4-5 orders of magnitude from $z\approx4$ to $z\approx0$ -- they are a predominantly early Universe phenomenon and therefore eluded discovery prior to JWST \citep[e.g.,][]{Park26, Ma26}. On the other hand, at higher redshift, $z\gtrsim8$ the number densities remain relatively flat even out to $z\approx10$, the highest redshift LRDs have been found so far \citep[e.g.,][]{Tanaka25}. 

For our fiducial forecast, we adopt the redshift-dependent LRD
abundance parameterization of \citet{Inayoshi2025}. He uses the LRD sample of \citet{kocevski2023hidden,kocevski2025rise} to measure the occurrence rate of LRDs as a function of cosmic time, and finds that the observed evolution is well described by a log-normal distribution (see the red solid line in the middle panel of Figure \ref{fig:threepanel}). This model captures the rapid emergence of LRDs at $z\sim 6$--$8$ and their sharp decline toward $z\lesssim 4$.

We convert this abundance model into the bin center number density which is plotted in the middle panel of Fig. \ref{fig:threepanel} alongside the DESI tracers (data from \citep[][]{desi2026data}), and reported in Table \ref{tab:forecast_inputs}. The resulting LRD densities are higher than  quasars at $z \gtrsim 3$, and remain relatively high across a wide redshift band. This modest evolution allows for wide bins in redshift, allowing for a large accessible comoving volume. 

\subsection{Survey Volume}
The comoving volume available to the survey follows from the footprint and the
redshift range. For a bin spanning $z_{\rm min}$ to $z_{\rm max}$,
\begin{equation}
  V_{\rm sur}(z_{\rm min}, z_{\rm max})
  \;=\;
  \frac{4\pi}{3}\,f_{\rm sky}\,
  \bigl[D_C^{\,3}(z_{\rm max}) - D_C^{\,3}(z_{\rm min})\bigr],
\end{equation}
where $D_C(z) = (c/H_0)\int_0^z dz'/E(z')$ is the line-of-sight
comoving distance in our fiducial flat $\Lambda$CDM cosmology, and
$f_{\rm sky}\simeq 0.34$ for a
DESI-like footprint.  Because the LRD abundance
stays high and only mildly evolving across $4\lesssim z\lesssim8$ (middle panel
of Fig.~\ref{fig:threepanel}), we integrate over wide $\Delta z = 1-2$ bins without
the density falloff that limits other tracers. We treat each bin as a single quasi-homogeneous sample, an assumption that larger samples of future data can test.   For the four LRD bins $ z=\{4{-}5,\,5{-}6,\,6{-}7,\,7{-}9\}$ we obtain $V_{\rm sur} = 153,\,137,\,122,\,209\,{\rm Gpc}^3$ respectively, shown in the lowest
panel of Fig.~\ref{fig:threepanel}.

\begin{table*}
\centering
\caption{%
  Little Red Dot BAO forecast inputs for all bin centers used in this analysis.
  The values for each of the DESI tracers are also tabulated for ease of
  comparison. DESI bias values and redshift ranges are from \cite{andrade2025validation}. DESI number densities are
  volume-averaged from the DR1 LSS catalogs \citep{desi2026data}. Survey volumes $V_\mathrm{sur}$ assume a 14,000\,deg$^2$
  footprint. The DESI $\sigma_{D_V/r_d}$ are from Table IV of \cite{DESI_DR2_II}.%
}
\label{tab:forecast_inputs}
\footnotesize
\renewcommand{\arraystretch}{1.15}
\setlength{\tabcolsep}{5.5pt}
\begin{tabular}{l c c c c c c}
\hline\hline
Bin
  & $z_\mathrm{center}$
  & $z$ range
  & $\bar{n}$\,[$h^{3}\,\mathrm{Mpc}^{-3}$]
  & $b$
  & $V_\mathrm{sur}$\,[Gpc$^3$]
  & $\sigma_{D_V/r_d}$\,[\%] \\
\hline
\multicolumn{6}{l}{\textit{DESI tracers}} 
  & \multicolumn{1}{r}{\textit{measured}} \\[1pt]
BGS           & 0.295 & $0.10$--$0.40$ & $5.2\times10^{-4}$ & 1.50 &   5.8 & 0.94 \\
LRG1          & 0.510 & $0.40$--$0.60$ & $5.3\times10^{-4}$ & 2.00 &  10.9 & 0.78 \\
LRG2          & 0.706 & $0.60$--$0.80$ & $5.3\times10^{-4}$ & 2.00 &  17.0 & 0.69 \\
LRG3$+$ELG1   & 0.934 & $0.80$--$1.10$ & $9.2\times10^{-4}$ & 1.60 &  35.0 & 0.46 \\
ELG2          & 1.321 & $1.10$--$1.60$ & $4.1\times10^{-4}$ & 1.20 &  75.7 & 0.72 \\
QSO           & 1.484 & $0.80$--$2.10$ & $3.1\times10^{-5}$ & 2.10 & 197.3 & 1.53 \\[3pt]
\multicolumn{6}{l}{\textit{LRD redshift bins}} 
  & \multicolumn{1}{r}{\textit{forecast}} \\[1pt]
LRD $4<z<5$   & 4.5 & $4.00$--$5.00$ & $6.3\times10^{-5}$ & 3.4 & 152.8 & 0.78 \\
LRD $5<z<6$   & 5.5 & $5.00$--$6.00$ & $1.3\times10^{-4}$ & 4.5 & 136.6 & 0.40 \\
LRD $6<z<7$   & 6.5 & $6.00$--$7.00$ & $1.6\times10^{-4}$ & 5.8 & 122.2 & 0.33 \\
LRD $7<z<9$   & 8.0 & $7.00$--$9.00$ & $1.1\times10^{-4}$ & 8.1 & 208.8 & 0.28 \\
\hline
\end{tabular}
\end{table*}
\begin{figure}[t!]
\hspace{-0.6cm}
\includegraphics[width=0.48\textwidth]{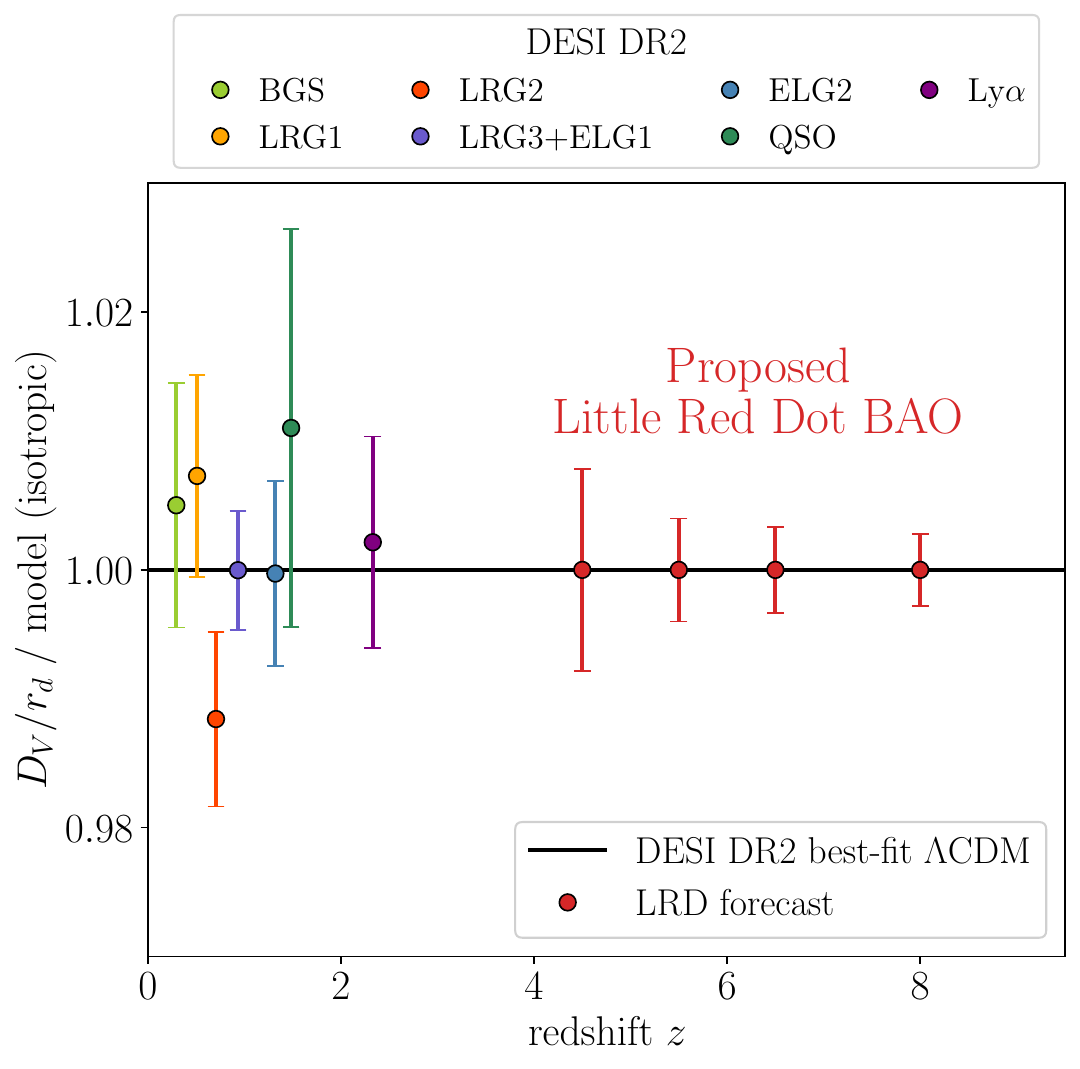}
\caption{Forecasted high-redshift BAO constraints from LRDs compared to
DESI DR2 isotropic BAO measurements from \cite{DESI_DR2_II}. Distances are shown as $D_V/r_d$
relative to the DESI DR2 best-fit $\Lambda$CDM model
\cite{DESI_DR2_II}, so the solid black line is unity by construction.
DESI DR2 measurements are shown at low redshift, while red points show the forecast LRD constraints for a DESI-footprint spectroscopic survey. LRD BAO would extend percent-level standard-ruler measurements to
$4\lesssim z\lesssim9$, providing a direct test of the matter-era
distance-scale extrapolation between the CMB and low-redshift tracers.}
\label{fig:forecast}
\end{figure}

\section{BAO Forecast}

We forecast the BAO constraining power of a spectroscopic survey with a DESI-like footprint using the LRD number density and bias presented in Figure \ref{fig:threepanel} and Table \ref{tab:forecast_inputs}. We assume a fiducial cosmology using Planck 2018 TT,TE,EE+lowE+lensing \cite[][]{Planck2018}. Forecasts are computed with the MultiFishLSS code \citep[][]{ebina2024cosmology}, which extends the FishLSS Fisher framework of \citet{sailer2021cosmology}.

The constraining power in each redshift bin is set primarily by the survey volume, which determines the number of accessible modes, and by the tracer number density and bias, which determine the significance with which these modes are measured. These three survey properties are summarized relative to DESI tracers in Fig.~\ref{fig:threepanel} and Table~\ref{tab:forecast_inputs}.

The Fisher matrix is evaluated for $\{\alpha_{\perp},\alpha_{\parallel}\}$, which can be interpreted as relative constraints on the BAO parameters $\{D_A(z)/r_d,\, r_d H(z)\}$ respectively, marginalizing over the linear bias and 15 broadband polynomial coefficients in each bin. Here, $D_A(z)$ is the angular-diameter distance and $H(z)$ is the Hubble expansion rate for each redshift bin.

\section{Results}

We report the forecast as a fractional constraint on $D_V/r_d$, where $r_d$
is the sound horizon at the drag epoch and
\begin{equation}
D_V(z)\equiv \left[zD_M^2(z)D_H(z)\right]^{1/3}
\end{equation}
Here $D_M(z)=(1+z)D_A(z)$ is the transverse comoving distance and
$D_H(z)=c/H(z)$ is the Hubble distance. 

Across the four bins, a DESI-footprint survey reaches
$\sigma_{D_V/r_d}=0.78,\,0.40,\,0.33,$ and $0.28\%$ at bin centers
$z=4.5,\,5.5,\,6.5,$ and $8.0$ (Fig.~\ref{fig:forecast},
Table~\ref{tab:forecast_inputs}). These are percent-level measurements in a currently unconstrained part of redshift space. They are competitive with DESI's low-redshift bins, and substantially smaller than the high-redshift quasar constraints. This follows from the combination of large volume, high
bias, and sufficient number density as seen in Figure \ref{fig:threepanel}. 

We stop the forecast at $z=9$, where the current empirical constraints on the
LRD population become increasingly uncertain and key rest-frame optical lines
move toward the edge of the most efficient JWST spectroscopic window. There is
currently no clear observational determination of the maximum redshift to which
the LRD population extends, or of how its abundance, bias, and spectroscopic
properties evolve at the highest redshifts. As larger samples become available,
it will be interesting to revisit whether the BAO lever arm can be extended to
even earlier times.

\section{Conclusion}

We have shown that LRDs have the abundance, bias, and redshift reach to serve as a high-redshift BAO tracer. A DESI-footprint spectroscopic survey
could measure $D_V/r_d$ to percent-level precision across $4\lesssim
z\lesssim9$, a range inaccessible to existing galaxy surveys.

Such a measurement would be an independent standard ruler between the CMB at early times and low-redshift tracers. This is the era in which dark energy is dynamically negligible, so an LRD BAO measurement would test the CMB-calibrated $\Lambda$CDM distance extrapolation before the beginning of late-time cosmic acceleration. 
It does not replace low-redshift BAO as a
probe of cosmic acceleration, but it adds a complementary and independent matter-era anchor between the CMB and the late-time Universe.

This measurement requires capabilities beyond those of current facilities. Spectroscopic redshifts for LRDs are most efficiently obtained via H$\alpha$, which at $z>3$ falls at $>2$ micron, beyond the sensitivity and wavelength coverage of current or planned wide-field facilities (e.g., Roman and Euclid spectroscopy are limited to $<2.2$ micron, SPHEREx probes to $5\mu$m, but does not have the necessary sensitivity). JWST has the correct wavelength coverage, but not the survey speed (its largest contiguous
extragalactic surveys span $<1\,{\rm deg}^2$ \citep[e.g.,][]{Casey23}, four orders of magnitude short
of the $\sim10^4\,{\rm deg}^2$ of a DESI-scale BAO survey). LRD BAO is
therefore not a JWST program but a science case for a future wide-field near-
to mid-infrared spectroscopic mission. Such a survey could be completed within a standard mission lifetime by, e.g., a probe-class telescope that pre-selects LRD targets in a brief initial imaging survey and follows with wide-area spectroscopy. If current abundance and clustering
estimates hold in larger samples, LRDs could provide the first
precision large-scale-structure survey deep in the matter-dominated era and an important benchmark in testing our standard cosmological model.

\begin{acknowledgments}
\noindent {\bf Acknowledgments}: The authors would like to thank Wayne Hu, Martin White, Bill Holzapfel, Haruki Ebina, Jorryt Matthee, and Anna de Graaff for useful conversations.  This document was prepared using the resources of
the Fermi National Accelerator Laboratory (Fermilab),
a U.S. Department of Energy, Office of Science, Office
of High Energy Physics HEP User Facility. Fermilab
is managed by Fermi Forward Discovery Group.
J.Z. is also supported by the Kavli Institute for Cosmological Physics.  This research used data obtained with the Dark Energy Spectroscopic Instrument (DESI). DESI construction and operations is managed by the Lawrence Berkeley National Laboratory. RPN acknowledges the generous support of Neil and Jane Pappalardo via the MIT Pappalardo Fellowship in Physics. RPN acknowledges funding from JWST-GO-5224. The authors used Anthropic Claude (including models through Fable 5) as interactive assistants for literature organization, manuscript and \LaTeX\ preparation, and coding and figure-workflow support.  The authors directed, checked, and revised all outputs used here.  All scientific judgments, derivations, calculations, interpretation, and responsibility for the manuscript remain solely with the authors. This material is based upon work supported by the U.S. Department of Energy, Office of Science, Office of High-Energy Physics, under Contract No. DE–AC02–05CH11231, and by the National Energy Research Scientific Computing Center, a DOE Office of Science User Facility under the same contract. Additional support for DESI was provided by the U.S. National Science Foundation (NSF), Division of Astronomical Sciences under Contract No. AST-0950945 to the NSF’s National Optical-Infrared Astronomy Research Laboratory; the Science and Technology Facilities Council of the United Kingdom; the Gordon and Betty Moore Foundation; the Heising-Simons Foundation; the French Alternative Energies and Atomic Energy Commission (CEA); the National Council of Humanities, Science and Technology of Mexico (CONAHCYT); the Ministry of Science and Innovation of Spain (MICINN), and by the DESI Member Institutions: www.desi.lbl.gov/collaborating-institutions. The DESI collaboration is honored to be permitted to conduct scientific research on I’oligam Du’ag (Kitt Peak), a mountain with particular significance to the Tohono O’odham Nation. Any opinions, findings, and conclusions or recommendations expressed in this material are those of the author(s) and do not necessarily reflect the views of the U.S. National Science Foundation, the U.S. Department of Energy, or any of the listed funding agencies.
\end{acknowledgments}

\bibliography{bibliography}

\end{document}